\documentstyle[12pt]{article}

\newcommand{\la}[1]{\label{#1}}
\newcommand{\beq}{\begin{equation}}
\newcommand{\eeq}{\end{equation}}
\newcommand{\ie}{{\em i.e.}}
\newcommand{\eg}{{\em e.g.}}

\newcommand{\eq}{Eq.~}
\newcommand{\eqs}{Eqs.~}
\newcommand{\fig}{Fig.~}

\newcommand{\sect}{Sec.~}
\newcommand{\simls}{\hbox{$\,$\raise.5ex\hbox{$<$}
       \kern-1.1em \lower.5ex\hbox{$\sim$}$\,$}}
\newcommand{\simgt}{\hbox{$\,$\raise.5ex\hbox{$>$}
       \kern-1.1em \lower.5ex\hbox{$\sim$}$\,$}}
\let\oref=\ref
\renewcommand{\ref}[1]{(\oref{#1})}

\newcommand{\T} {\hat T}
\newcommand{\Tf} {\hat T_f}
\newcommand{\vdet} {v_{\rm det}}
\newcommand{\vdef} {v_{\rm defl}}
\newcommand{\vsh} {v_{\rm sh}}
\newcommand{\vin} {v_{\rm in}}
\newcommand{\vout} {v_{\rm out}}
\newcommand{\eph} {\epsilon_h}
\newcommand{\epq} {\epsilon_q}
\newcommand{\TqD} {T_q({\rm D})}
\newcommand{\TqJ} {T_q({\rm J})}
\newcommand{\TfG} {T_f({\rm G})}
\makeatletter
\def\fnum@figure{Fig.~\thefigure}
\makeatother

\begin{document}
\pagestyle{empty}
\setlength{\parindent}{0.0cm}

\begin{center}
\vspace*{1.5cm}
{\Large\bf LARGE SCALE INHOMOGENEITIES FROM} \\
\vspace*{2mm}
{\Large\bf THE QCD PHASE TRANSITION} \\
\vspace*{6mm}
{\large\bf J.~Ignatius$^{a}$, K.~Kajantie$^{b}$,
H.~Kurki-Suonio$^{a}$ and M.~Laine$^{b}$} \\
\vspace*{3mm}
{\sl  $\mbox{}^{a}$Research Institute for Theoretical Physics and \\
$\mbox{}^{b}$Department of Theoretical Physics, }
\\
{\sl  P.O.~Box 9, FIN-00014 University of Helsinki, Finland}
\\
\vspace*{3mm}
19 May 1994
\end{center}

\vspace*{-9cm}
\hfill Preprint HU-TFT-94-14
\vspace*{9cm}

\begin{center}
{\large\bf Abstract}
\end{center}
\vspace*{2mm}
We examine the first-order cosmological QCD phase transition for a
large class of parameter values, previously considered unlikely.
We find that the
hadron bubbles can nucleate at very large distance scales, they can
grow as detonations as well as deflagrations,
and that the phase transition may be completed without
reheating to the critical temperature.
For a subset of the parameter values studied, the
inhomogeneities generated at the QCD phase transition might have a
noticeable effect on nucleosynthesis.
\vspace*{3mm}


\vfill

\noindent \underline{ \mbox{ \hspace*{6.5cm} } }  \\
Electronic addresses: ignatius@phcu.helsinki.fi,
kajantie@phcu.helsinki.fi,\\
hkurkisu@pcu.helsinki.fi, mlaine@phcu.helsinki.fi

\newpage
\pagestyle{plain}
\setcounter{page}{1}
\setcounter{footnote}{0}
\setlength{\parindent}{0.85cm}

\section{Introduction}
\la{Intro}

Much of the interest in the cosmological first-order QCD phase
transition has been due to the possibility of affecting the
big-bang nucleosynthesis~\cite{AHS}.
This would require creating significant inhomogeneities in the baryon
number density, with a characteristic distance scale of the order of
$10^{-3} t_H$.  The typical distance $l_n$ between bubbles
nucleated in
the phase transition is related to the bubble surface tension, $l_n
\propto \sigma^{3/2}$.
The excitement has diminished as lattice calculations have yielded
rather low estimates of $\sigma$, indicating much smaller distance
scales.

The purpose of this paper is to point out that a low $\sigma$
does not necessarily imply short distances. The reason
is that in addition to $\sigma$, the distance scale depends on
the latent heat $L$. As the phase transition appears to be at
most only weakly first order, surface tension and latent heat should
both be small compared to energy scales at the transition.
Using the bag equation of state and classical nucleation theory,
we find that even for small $\sigma$,
there is a large parameter space where the
phase transition is preceded by considerable supercooling, and
the distances between critical bubbles are large. Interestingly,
we also find that with a more realistic equation of state
than that of the bag model, the distances get even larger. It will
be very enlightening to redo our calculations, when the
whole equation of state of the QCD matter can be extracted
from lattice calculations.

In view of nucleosynthesis, not only the distances between critical
bubbles but also the later stages of the phase transition are
important.
Since we do not know the microscopic physics operating at
the phase transition front, we cannot give a definite solution
to the problem. However, we can
analyze what processes are allowed by hydrodynamics.
We find that the hadron bubbles could grow as detonations and that
there is not necessarily a stage of slow growth in thermodynamical
equilibrium during the phase transition. This leads to a rather
unconventional picture of the phase transition. We also find that if
this picture is correct, then it is unlikely that the QCD phase
transition has a noticeable effect on nucleosynthesis, in spite
of the large distance scales. On the other hand, it is also possible
that
the hadron bubbles grow as deflagrations, the distance scales are
equally large as above, and the QCD phase transition may have
a noticeable effect
on nucleosynthesis. Clearly, for conclusive results,
a better knowledge of the parameter values must be acquired.

The plan of the paper is the following.
In \sect\oref{Nucl} we review classical nucleation theory and
calculate
the nucleation temperature and the distances between critical bubbles
in the QCD phase transition, using the bag equation of state.
In \sect\oref{EOS} we discuss the effects of a more realistic
equation of state. Sec.~\oref{Detodefl} contains a hydrodynamical
analysis of the possible growth mechanisms of the hadron bubbles, and
in \sect\oref{Baryon} we discuss the effect of the QCD phase
transition on nucleosynthesis. The conclusions are in
\sect\oref{Concl}.

Since the hydrodynamics of bubble growth in this context
has sometimes been analyzed
using a plane-symmetric geometry (\ie, 1+1 dimensions), we stress that
in this paper we use spherical geometry everywhere.

\section{Nucleation}
\la{Nucl}

We review the classical thermal nucleation of thin-wall bubbles
with the bag equation of state
\beq
   p_q = a_q T^4 - {L\over4}, \qquad\qquad p_h = a_h T^4,
\la{eos}
\eeq
where $L = 4(a_q-a_h)T_c^4$ is the latent heat.
We use the notation $\T \equiv T/T_c$, where $T_c$ is the critical
temperature of the phase transition.
The radius of a critical bubble at temperature $\T < 1$ is
\beq
   r_c = {2\sigma\over p_h-p_q} = {8\sigma \over L}{1\over 1-\T^4},
\la{rc}
\eeq
where $\sigma$ is the bubble surface tension.
The free energy expended to nucleate a critical bubble is
\beq
   W(r_c) = {4\pi\over3}\sigma r_c^2 = {16\pi\over3} {\sigma^3
   \over (p_h-p_q)^2}.
\la{Wrc}
\eeq
The nucleation rate is
\beq
   p(t) = CT_c^4e^{-S(t)},
\la{pt}
\eeq
where
\beq
   S(t) \equiv {W(r_c)/T} = {256\pi\over3} {\sigma^3\over L^2T_c}
   {1\over\T}{1\over(1-\T^4)^2},
\la{St}
\eeq
and $C$ is a prefactor roughly of order unity.

There are thus two essential parameters, $L$ and $\sigma$.  The latent
heat $L$  gives the difference of $a_q$ and $a_h$.  Otherwise the
values
of $a_q$ and $a_h$ affect the nucleation only slightly through the
expansion rate of the Universe.  We are interested in the case of a
small $L$.  Therefore, $a_q$ and $a_h$  cannot be taken to have their
ideal gas values~--- $51.25\pi^2/90$ and $17.25\pi^2/90$, which
correspond to
a large latent heat $L = 14.9T_c^4$~--- but must lie closer to each
other
somewhere between these values.  In reality they are functions of
temperature $a_q(T)$, $a_h(T)$, but until \sect\oref{EOS},
we take them to be constant.

With constant $a_q$ and small $L$,
the expansion (cooling) timescale is approximated by
\beq
   tT^2 = {\rm const.} = t_c T_c^2.
\la{tT}
\eeq
The Universe reaches the critical temperature at $t = t_c$.
The Universe then supercools to $T<T_c$ and the nucleation rate
begins to increase rapidly.  A nucleated bubble grows by detonation or
deflagration, preventing further nucleation inside the volume
$(4\pi/3)v^3(t-t')^3$, where $t'$ is the nucleation time.
In the case of
detonation the relevant velocity $v$ is that of the phase boundary
(detonation front), $v = \vdet$.  In the case of deflagration, it is
that of the shock driven ahead of the phase boundary, $v = \vsh$.
In both cases, $c_s < v < 1$, where $c_s = 1/\sqrt{3}$ is the speed of
sound.  The expansion rate of the Universe can be ignored during this
short period of rapid bubble growth.  The fraction of space affected
is then~\cite{GuthWei}
\beq
   F(t) = 1 - \exp\biggl[-\int_{t_c}^t dt'p(t')
{4\pi\over3}v^3(t-t')^3\biggr].
\la{Ft}
\eeq
This grows rapidly from $F\simeq0$ to $F\simeq1$ at the phase
transition
time $t_f$, which we define by
\beq
   F(t_f) \equiv 1 - 1/e.
\la{tf}
\eeq
Bubble nucleation then ceases.

Let us note that for very
slow deflagrations, $\vdef \simls 0.1$, the shock preceding the
deflagration front is extremely weak.  Between the shock and the
deflagration front, the temperature increases
continuously~\cite{Kurki-Suonio}. However, even
the temperature at the deflagration
front, $T_q$, may have been raised so little above the nucleation
temperature, $T_f$, that the
nucleation of new bubbles is not appreciably suppressed.
Thus the velocity $v$ should be
the deflagration velocity $\vdef$, instead of the shock velocity
$\vsh$.
As will be seen below, this leads to much shorter distance scales.

Approximating
\beq
   S(t) \simeq S(t_f) - S'(t_f)(t_f-t)
\la{Sapx}
\eeq
and extending the integral to $-\infty$, \eq\ref{tf} becomes
\beq
   S'(t_f)^4 = 8\pi v^3CT_c^4e^{-S(t_f)}.
\la{Seq}
\eeq
The typical distance $l_n \equiv n^{-1/3}$ between centers of
neighboring bubbles is obtained from their number density
\beq
   n = \int dt' p(t') \bigl[1-F(t')\bigr]
   \simeq CT_c^4 {e^{-S(t_f)} \over |S'(t_f)|}
   = {|S'(t_f)|^3 \over 8\pi v^3}.
\la{N}
\eeq
The above approximations require
\beq
   {|S''|\over|S'|^2} \ll 1
\la{apx1}
\eeq
and
\beq
   |S'|(t_f-t_c) \gg 1.
\la{apx2}
\eeq

{}From \eqs\ref{St} and \ref{tT} we obtain
\beq
   S(t_f) = {A\over\Tf y^2},
\la{S0}
\eeq
\beq
   S'(t_f) = -{A\over t_c} {\Tf(8-9y)\over 2y^3},
\la{S1}
\eeq
and
\beq
   S''(t_f) = {A\over t_c^2} {\Tf^3\over 4y^4} (63\Tf^8+34\Tf^4-1),
\la{S2}
\eeq
where
\beq
   A \equiv {256\pi\over3} {\sigma^3\over L^2T_c}
\la{A}
\eeq
and
\beq
   y \equiv 1-\Tf^4.
\la{y}
\eeq
The nucleation time $t_f$ and temperature $T_f$ are solved from
\eq\ref{Seq}.  Taking the logarithm and using \eqs\ref{S0}
and \ref{tT}
it becomes
\beq
   y^2 = {A\over\Tf S(t_f)} = {A\over\Tf\bigl[S_A -
   4\ln|S'(t_f)t_c|\bigr]},
\la{y2}
\eeq
where $S_A \equiv \ln 8\pi v^3CT_c^4t_c^4 \simeq 4\ln T_c t_c \simeq
170$. This gives an equation for $y$,
\beq
   y = {A^{1/2}\over(1-y)^{1/8}
    \Bigl[S_A-4\ln|A(1-y)^{1/4}(8-9y)/(2y^3)|
   \Bigr]^{1/2}},
\la{yeq}
\eeq
which can be solved iteratively.
The phase transition temperature is then
\beq
   T_f = (1-y)^{1/4}T_c
\la{Tf}
\eeq
and the typical bubble distance
\beq
   l_n = \pi^{1/3} {v\over A} {2y^3\over(1-y)^{1/4}(8-9y)} t_H,
\la{ln}
\eeq
where $t_H = 2t_c$ is the Hubble distance at $t = t_c$.
We plot these quantities in \fig\oref{figB}.  They depend on the
parameters $\sigma$ and $L$ only through the combination \ref{A}.

The approximations \ref{apx1} and \ref{apx2} fail near
$\sigma^3/L^2T_c
= 0.25$ when the supercooling $1-\Tf$ and the distance scale $l_n$ are
becoming very large.  The nucleation rate, according to \eq\ref{St},
is beginning to grow more slowly and has a maximum
at $T = T_c/\sqrt3$.
This would indeed lead to extremely deep supercooling and very large
distances. Of course, our equation of state \ref{eos},
tuned to be valid near $T_c$, is then no longer applicable.

The critical radius
\beq
   r_c = {2\sigma\over p_h-p_q} = {8\sigma\over Ly}
\la{rceq}
\eeq
must also not be too small ($\sim T_c^{-1}$) for the above thin-wall
nucleation to apply.
{}From \fig\oref{figC} we see that for small $\sigma$, the critical
radius
is reasonably large, and almost independent of $L$.


The values of $L$ and $\sigma$ indicated by
pure glue lattice Monte Carlo simulations are of the order
$L\approx2T_c^4$,
$\sigma\approx0.02T_c^3$~\cite{qcdpax,grossmann,iwasaki}.
However, the uncertainty in these values is very large and, in
particular, there is no lower limit.  Thus, based on present
knowledge,
either one could be arbitrarily small. {}From \fig\oref{figC} we see
that,
\eg, length scales of $l_n/vt_H\approx 10^{-3}$--$10^{-2}$
are possible.
This is a distance scale large enough to affect nucleosynthesis.
The corresponding critical radii $r_c$ are large enough so that
the thin-wall
calculation should be valid.

\section{The effect of the equation of state}
\la{EOS}

Lattice Monte Carlo simulations imply that the energy density must
have a very  strong variation within a narrow temperature interval
($\leq$~10~MeV)  in QCD with physical quark masses~\cite{Petersson}.
Combined with the smallness of the latent heat this means that  the
realistic equation of state must differ even qualitatively from that
of  the bag model. This is illustrated in \fig\oref{fig:3EOS}.  As has
been discussed above, the parameter values of the naive bag model can
be  corrected to reproduce a desired latent heat. However, this
corrected bag model (``bag model of text'') cannot mimic at all the
strong variation in the energy density. In this section we will
analyze how the large derivative of the energy density curve, \ie,~the
large heat capacity,  affects the nucleation.

In the case of the nucleation calculation we may assume that
the Universe is,
at least locally, in the quark phase and very near thermal
equilibrium.
Furthermore, we can ignore the tiny quark and lepton chemical
potentials.
We denote the cosmic scale factor of the Robertson--Walker metric
by $R$, the energy density\footnote{In this section, all quantities
are measured in the supercooling quark phase.}
by $\epsilon$, and the entropy density by $s$.
{}From the equation ${\rm d}(\epsilon R^3)=-p{\rm d}(R^3)$
it follows that
\beq
  \frac{{\rm d}R}{R} = - \frac{{\rm d}\epsilon}{3(\epsilon +p)}=
-\frac{c(T)}{3s(T)} \frac{{\rm d}T}{T} ,
\la{dS2}
\eeq
which can also be written as
the equation ${\rm d}(sR^3)=0$ for
entropy conservation.
Here $c(T)$ is the density of heat capacity (specific heat) of the
quark phase
in constant volume. It satisfies the relations
$c(T) \!=\! {\rm d} \epsilon / {\rm d} T \!=\!
T ({\rm d}s / {\rm d}T)
 \!=\! T({\rm d}^2p / {\rm d}^2T)$, where $p$ is the pressure.
Equation~\ref{dS2} tells how the relation between expansion
and cooling depends on the equation of state.
Notice that in the case of the bag equation of state the factor
$\: c/3s \:$ equals unity, and the simple relation
$\: T \propto 1/R \:$ is valid.

The expansion rate of the Universe is determined from the Friedmann
equation,
\beq
  \frac{ \dot{R}^2 }{R^2} = \frac{8 \pi \epsilon(T)}{3 M_{\rm Pl}^2} ,
\la{Friedmann}
\eeq
where $M_{\rm Pl}$ denotes the Planck mass. The vanishingly small
curvature
term was omitted.
Eliminating $\: {\rm d}R / R \:$ from \eqs\ref{dS2} and
\ref{Friedmann}
gives the cooling rate of the Universe:
\beq
  {\rm d}T = - \frac{3s(T)}{c(T)} \times
    \sqrt{ \frac{8 \pi \epsilon(T)}{3} }
    \frac{T}{M_{\rm Pl}} \, {\rm d}t .
\la{dTdt}
\eeq
If the energy density curve $\epsilon(T)$ is very steep, the first
factor is
much smaller than unity. The Universe expands and the energy density
decreases,
but this causes only very slow cooling. In other words,
the total energy
density determines by \eq\ref{Friedmann} the expansion rate
of the Universe,
and this in turn determines by \eq\ref{dS2} the value of
the derivative $\dot{\epsilon}(t)$;
but then ${\rm d}T/{\rm d}t=\dot{\epsilon}(t)/c(T)$ is very small
for large heat capacity.

Since the Universe cools slowly, it has time to nucleate
even at a large action, and therefore the true
nucleation temperature is slightly
higher than that of the bag model. To see this, notice first that
compared with the time spent in the supercooled state,
the period of time during which essentially all the nucleation
takes place
is short, and the cooling rate~\ref{dTdt} is practically constant
during
this period. We define the parameter $\delta$,
\beq
  \delta = \frac{c(T_f)}{3s(T_f)} ,
\la{delta}
\eeq
where $T_f$ is the temperature at the nucleation time $t_f$.
In Fig.~\oref{fig:3EOS}, $\delta$ is roughly the ratio of slopes of
the  realistic equation of state, and the bag equation of state.
The nucleation temperature is determined from \eq\ref{Seq}.
However, now the pressure difference
$\: p_h(T) - p_q(T) \:$ is unknown, and thus is also the nucleation
action~$S(T)$. Still, we can find out what is the effect of $\delta$:
in the logarithm of \eq\ref{Seq} the value
of the constant $\: S_A \!\simeq\! 170 \:$ increases
by $\: 4 \log \delta \:$.
As long as this increase is small compared with 170
the amount of supercooling, defined as $\: 1 - T_f/T_c $,
decreases only little.

Although a large value of $\delta$ barely changes $T_f$, it has
another significant effect. Physically the most important
quantity related to nucleation is the average
distance between nucleation centers immediately after the phase
transition, $l_n$, defined between \eqs\ref{Seq} and~\ref{N}.
Its dependence on $\delta$ is seen from \eq\ref{N} to be
\beq
  l_n \: \propto \: \delta ,
\la{ldelta}
\eeq
where the small correction coming from the change in $T_f$ was
left out.
This result tells that a steep drop in the energy density
$\epsilon(T)$ increases significantly the typical distance
between bubbles.
This happens because when the cooling rate is lower it takes
more time
for the nucleation action $S(T)$ to decrease by a certain amount.

To get an order of magnitude estimate for $\delta$, we assume  that
the energy density changes from its asymptotic value in the quark
phase to its asymptotic value in the hadron phase within $10$ MeV.
This is an upper limit  indicated by lattice
calculations~\cite{Petersson}. Then we get $\delta\approx 8$.  Thus
the distance scales would be an order of magnitude larger than those
shown in \fig\oref{figC}. If the true temperature interval for the
rapid change in energy density were smaller, say $1$ MeV, then
distance scales would be correspondingly larger, $\delta\approx75$.

Finally, we note that the validity conditions of the nucleation
calculation, \eqs\ref{apx1} and \ref{apx2}, do not depend in practice
on $\delta$ unless the specific heat of the quark phase, $c(T)$,
changes rapidly within the temperature interval $\: T_c \ldots T_f$.

\section{Detonations and deflagrations}
\la{Detodefl}

A bubble of $h$ phase surrounded by supercooled $q$ phase has two
modes of growth available, detonations and
deflagrations~\cite{LL,Courant,GKKM}. In a deflagration bubble the
fluid inside the bubble is at rest, but the growing bubble is
surrounded by a shock wave moving out ahead of the phase transition
(deflagration) front.  In a detonation bubble the phase transition
(detonation) front advances into the fluid which is at rest, but is
followed by a rarefaction wave where the fluid flows outwards
following the detonation front.  Relativistic detonation bubbles have
been discussed in Ref.~\cite{Steinhardt}, and relativistic
deflagration bubbles in Ref.~\cite{Kurki-Suonio}.

Consider fluid flow through the phase transition front in the rest
frame of the front.  The inflow is subsonic for a deflagration,
$\vin<c_s$, but supersonic for a detonation, $\vin>c_s$.  These
processes are further divided into weak and strong depending on the
outflow velocity. For weak processes the nature of the flow velocity
does not change, \ie, $\vout<c_s$ for weak deflagrations, and
$\vout>c_s$ for weak detonations.  For strong processes it changes,
\ie, $\vout > c_s$ for strong deflagrations, and $\vout<c_s$ for
strong detonations.  The case where $\vout = c_s$ is called a Jouguet
process.

In the rest frame of the unaffected fluid,
all detonations and strong
deflagrations move faster than sound, whereas weak deflagrations are
subsonic.

Because of restrictions on the fluid flow pattern from the bubble
geometry, strong detonations are not possible for phase transition
bubbles~\cite{Steinhardt}.  (This is true also for 1-dimensional
``bubbles'').

In classical combustion theory for chemical burning the internal
structure of the combustion front rules out strong deflagrations and
weak detonations~\cite{LL,Courant}. In particular, the internal
structure of a detonation front consists of a shock heating up the
medium to initiate combustion, immediately followed by a deflagration.
For a weak detonation, this deflagration would be a strong one.  Thus
the impossibility of strong deflagrations implies the impossibility of
weak detonations.

The internal structure of a phase transition front is different from a
combustion front.  Heating by a shock does not facilitate the phase
transition, and the structure of a detonation front is not
shock$+$deflagration.  Therefore weak detonations are not ruled
out~\cite{Laine}.  Thus slowly growing bubbles ($v < c_s$) will be
weak deflagrations and fast ones ($v>c_s$) will probably be weak
detonations~\cite{IKKL}.  Strong deflagrations might also  be possible
in some cases.

We denote by $T_h$ and $T_q$ the temperatures of the two phases at the
phase transition front. For a detonation bubble, $T_q$ will be the
phase transition temperature derived in \sect\oref{Nucl}, $T_q = T_f$,
but $T_h>T_f$, and we can even have $T_h>T_c$.  The rarefaction wave
cools the $h$ phase, so the final temperature will be below $T_h$ (and
$T_c$). For a deflagration bubble, $T_h$ will be the final temperature
of the $h$ phase, whereas $T_q$ is not the initial temperature $T_f$,
but  $T_q>T_f$ (and could exceed $T_c$), as the $q$ phase has been
heated by the shock wave.

The detonation and deflagration solutions are obtained from the
hydrodynamical conditions of energy and momentum conservation. These
processes must also satisfy the condition of non-negative entropy
production~\cite{LL,Courant,GKKM}.  These constraints do not fix the
process uniquely for a given initial temperature  $T_f$.  Instead, we
have a one-dimensional family of allowed solutions for each $T_f$,
with different temperatures $(T_h,T_q)$ and different bubble growth
velocities $\vdef$ or $\vdet$.  This family may contain both
deflagrations and detonations. Weak deflagrations are allowed for any
$T_f<T_c$, but detonations and strong deflagrations require a minimum
amount of supercooling, see \fig\oref{eheqfig}.

Below we give exact results for the bag equation of state \ref{eos}.
If the energy densities are scaled by the bag constant $B = L/4 =
(a_q-a_h)T_c^4$, and the temperatures by $T_c$, these results can be
given in terms of a single parameter
\beq
   r \equiv {a_q\over a_h} =
    \biggl[1-{L\over4a_qT_c^4}\biggr]^{-1} > 1.
\la{r}
\eeq
We identify a process by a point on the $(\eph/B,\epq/B)$-plane,
where
$\eph$ and $\epq$ are the energy densities of the two phases at
the phase
transition front.  There are a number of special points
in this plane
(\fig\oref{eheqfig}).

Point C corresponds to $T_h = T_q = T_c$.  This is the limit of weak
deflagrations as $T_f \rightarrow T_c$.   It is at the
$\vdef \rightarrow 0$
limit (the diagonal line through C).
Point D is the (weak) detonation which requires the least
supercooling.
This point is at the $\vdet \rightarrow 1$ limit (the diagonal line
through D).  The velocities change steeply near these diagonals.
Points J and G are the Jouguet detonations and deflagrations,
respectively, requiring the least supercooling.  The coordinates
$(\eph/B,\epq/B)$ of these points are
\beq
   C = \biggl({3\over r-1}, {4r-1\over r-1}\biggr),
\la{C}
\eeq
\beq
   D = \biggl({3r\over r-1}, {r+2\over r-1}\biggr),
\la{D}
\eeq
\beq
   J = \bigl(s_J+t_J, s_J-t_J),
\la{J}
\eeq
\beq
   G = \bigl(s_G+t_G, s_G-t_G),
\la{G}
\eeq
where
\beq
   s_J = {2r\over r-1}(1+\cos2\beta),
\la{sJ}
\eeq
\beq
   t_J = 2\sqrt{2r\over r-1}\cos\beta
\la{tJ}
\eeq
\beq
   s_G = {r\over r-1} (1 + 2\sin^2\beta +
   2\sqrt{3}\cos\beta\sin\beta),
\la{sG}
\eeq
\beq
   t_G = -\sqrt{2r\over r-1} (\cos\beta + \sqrt{3}\sin\beta),
\la{tG}
\eeq
and
\beq
   \beta \equiv {1\over3}\arctan\sqrt{r+1\over r-1}.
\la{beta}
\eeq

Converting the energy densities to temperatures by
\beq
   T_h^4 = {r-1\over3}{\eph\over B}T_c^4,\;\;\; T_q^4 =
   {r-1\over 3r}{\epq-B
   \over B}T_c^4,
\la{etoT}
\eeq
we have the following results.
The maximum temperature for which the weak detonations are allowed is
\beq
   T_f = \TqD = r^{-1/4}T_c.
\la{TqD}
\eeq
At the detonation front we then have  $T_q = r^{-1/4}T_c$, $T_h
  = r^{1/4}T_c$.
The maximum temperature for which Jouguet detonations are allowed, is
\beq
   T_f = \TqJ = \biggl[{r+1\over 3r} + {2\over3}\cos2\beta
   - {2\over3}\sqrt{2}\sqrt{r-1\over r}\cos\beta\biggr]^{1/4}T_c.
\la{TqJ}
\eeq
We always have $\TqJ < \TqD < T_c$.

For deflagration bubbles $T_q\neq T_f$.  To relate these two
temperatures we have numerically integrated the flow equations for the
region between the shock and the deflagration
fronts~\cite{Kurki-Suonio}.
We denote the maximum temperature at which strong
deflagrations are allowed by $\TfG$.  The temperatures $\TqD$, $\TqJ$,
and $\TfG$ are plotted as a function of $r$ in \fig\oref{tfg}.

If the Universe supercools very much, the phase transition is
not able to
reheat it back to $T_c$.  The limiting supercooling temperature $T_r$
is obtained from
\beq
   \epq(T_r) = \eph(T_c).
\la{Trdef}
\eeq
For the bag equation of state \ref{eos}, this gives
\beq
   T_r = \biggl({4-r\over3r}\biggr)^{1/4}T_c
       = \biggl({1\over r} - {r-1\over3r}\biggr)^{1/4}T_c.
\la{Trbag}
\eeq

If $T_f>T_r$, the phase transition reheats the Universe to $T_c$
before completing.  The rapid detonation/deflagration stage is
followed by a slower stage, where both phases coexist in thermal
equilibrium at $T_c$, and the phase transition proceeds only as the
Universe expands~\cite{KajKur}. If $T_f<T_r$, the
detonation/deflagration takes the transition to completion.  If $r>4$,
\ie, $B>\eph(T_c)$, then reheating to $T_c$ is possible from
arbitrarily low temperatures.

Comparing \eq\ref{TqD} with \eq\ref{Trbag} we note that $T_r$ is
always below $\TqD$, although for small $r$, these are close to each
other. Thus in those cases where the Universe has supercooled so much
that it will not reheat back to $T_c$, weak detonations are always
allowed.  The similar statement for Jouguet detonations becomes true
for $r > 1.644$.

In \sect\oref{Nucl} we related $T_f$ to $\sigma$ and $L$.  The
parameter $r$ depends on $a_q$ in addition to $L$.  By making some
assumption about $a_q$, we can convert $\sigma$ and $L$ to $T_f$ and
$r$, and classify points in the $(L,\sigma)$ parameter space according
to which processes are allowed. This is done in \fig\oref{siglnew}.

As a concrete example,  let us inspect the case $\sigma \!=\! 0.01$.
Now it is seen from  \fig\oref{siglnew} that if $L \!=\! 1$, only weak
deflagrations are allowed and the Universe does not reheat to the
critical temperature. If $L \!=\! 0.1$, weak deflagrations are
possible as well, and  the Universe reheats back to $T_c$. If $L \!=\!
0.01$ even Jouguet processes and strong deflagrations are allowed.

To summarize the results of this section,  for a given phase
transition temperature $T_f<T_c$, there is a one-dimensional family of
allowed bubble growth processes. This family will always include weak
deflagrations.  It may also include weak detonations (if $T_f <
\TqD$), Jouguet detonations (if $T_f \le \TqJ$), Jouguet deflagrations
(if $T_f \le \TfG$), and strong deflagrations (if $T_f < \TfG$).
Which of these allowed processes actually occurs, depends on the
dissipative mechanisms internal to the front which determine the
propagation speed of the transition~\cite{LL,Courant,IKKL}. Even
though external conditions may allow detonations, the actual process
could still be a slowly propagating weak deflagration.

\section{Baryon number}
\la{Baryon}

Much of the interest in the QCD phase transition in cosmology stems
from the possibility of leaving behind strong inhomogeneities in the
baryon number, and maybe thus affecting big bang nucleosynthesis.  The
baryon number in the $q$ phase is carried by massless quarks, but in
the $h$ phase it is carried by nucleons, with $m \gg T_c$.  The baryon
number does not penetrate the phase boundary easily, and accumulates
as a layer on the $q$ side of the phase boundary.  As more baryon
number accumulates onto this layer, more will also leak through, but
the net effect is that of dragging baryons towards the regions which
remained longest in the $q$
phase~\cite{FMA,Kurki-SuonioB,KapOli,SKAM}.

Even assuming we know the hydrodynamic details of the phase transition
discussed above, it is difficult to estimate the shape and density
contrast of the baryon number inhomogeneity.   It depends on the rate
of baryon transport within each phase and baryon penetration of the
boundary, which are not known.

To have a significant effect on nucleosynthesis, a number of
conditions need to be satisfied.  1) The distance scale should be
large, $l_n > 1\;{\rm m}\approx10^{-4}t_H$.  2) The high-density
regions should contain most of the total baryon number.  3) The
density contrast $R$, \ie,~the ratio of baryon number density in the
high-density region to that  in the low-density region, must be large,
$R_{\rm final} > (n_p/n_n)_{\rm nucleosynthesis} \sim 7$.

We have argued in Sections~\oref{Nucl} and~\oref{EOS} that condition
1 could be satisfied. Because of the difficulty of getting baryon
number through the phase boundary, condition 3 does not appear
unreasonable.  Condition 2 is perhaps the most
difficult~\cite{Kurki-SuonioB}.

During the phase transition, baryon number has been collected onto a
layer on the surface of the bubble, with some thickness $d$.   The
baryon density in this layer is $R_{\rm layer}$ times larger than
elsewhere.  To have most of the baryon number in this layer, we must
have roughly  $R_{\rm layer} > l_n/d$.  If the thickness of the layer
is due to microscopic diffusion of baryon number away from the
boundary (in the $q$ phase), $d$ will be very small.  It has been
argued that turbulent transport will be much more effective than
microscopic diffusion, and lead to a much thicker layer, so that
condition 2 might be satisfied~\cite{MMAF}.

In the usual picture of this phase transition, most of the growth of
the $h$ regions will happen in the equilibrium stage of the
transition, very slowly.  In the picture we have presented here, which
leads to large distance scales $l_n$ even with a small surface tension
$\sigma$, this stage does not usually exist.  The phase transition is
completed by detonation or deflagration.

Especially, if the bubble growth process is a detonation, the phase
transition is completed rapidly.  The detonation bubbles grow with
$\vdet > c_s$.  In those regions where the detonation bubbles collide,
the transition is then already completed, and the turbulence caused by
the collision will operate only in the $h$ phase. Where there is any
$q$ phase left, the detonation front will move on  unaffected, and
once they have covered the space between them, the transition is over.
The baryon number accumulating on the phase boundary cannot escape
from the supersonically moving front.  Thus the layer should get no
thicker than a few fm.  This appears to make condition 2 impossible to
satisfy.

If the bubble growth process is a deflagration, the phase boundary
will be slower and turbulent baryon transport may be effective in
making the layers thicker. As mentioned in \sect\oref{Nucl},  the
distances between nucleated bubbles do not depend  on whether the
bubbles grow as deflagrations or detonations, unless the deflagrations
are exceedingly slow. Therefore, slowly growing deflagration bubbles,
with efficient turbulent baryon transport, seem to be able to affect
nucleosynthesis.

If the hadron bubbles grow as deflagrations, but with an exceedingly
low velocity, there may be another mechanism, in addition to
turbulence, for enhancing the baryon number. Neutrinos, which do not
carry any baryon number, may carry  a considerable part of the energy
flux.  The reason is that the hydrodynamical flux $w\gamma^2v$,
measured in the rest frame of the phase transition front, vanishes as
$v\to 0$. This curve is given as the solution of the equation
$p_q(T_q)=p_h(T_h)$. The solution does not agree with the curve
$T_q=T_h$, so the neutrino flux $(g_{*n} \, \pi^2 / 120) (T_q^4 -
T_h^4)$, where $g_{*n}$ denotes  the effective number of active
neutrino degrees of freedom,  remains non-zero. But as noted above,
for these very slow deflagrations the distance scales are also small.
However, according to \sect\oref{EOS} an extremely large  heat
capacity would increase the distance scale with several  orders of
magnitude. There is also another escape: suppose that the initial
growth  mechanism is a deflagration which is not very slow,  so that
distance scales are large, but that then  the Universe reheats to
$T_c$, where a stage of very slow  growth near thermodynamical
equilibrium follows. Here, neutrino transport may be
effective~\cite{AH}, in addition to turbulence. If this stage lasts
long enough, the baryon number  remaining at the high-density regions
could be enhanced. {}From Figs.~\oref{figC} and~\oref{siglnew} one can
see that this scenario is possible in a very small but non-vanishing
region of the parameter space, near \eg~the point $L\approx2T_c^4$,
$\sigma\approx0.3T_c^3$.

Finally, let us note that so far we have only studied  which
hydrodynamical processes are in principle possible, without being able
to fix a definite growth velocity. In Ref.~\cite{IKKL} a model is
presented which tries to fix this velocity, by introducing a
phenomenological dissipative constant $\Gamma$. With a dimensional
estimate $\Gamma\approx1\;T_c^{-1}$, and  the parameters
$L\approx0.1T_c^4$, $\sigma\approx0.1T_c^3$,  the bubbles grow as
deflagrations, with a velocity $v_{\rm defl}=0.1$. The neutrino flux
is vanishingly small, but turbulence might be effective.

\section{Conclusions}
\la{Concl}

Parametrizing the QCD phase transition with the latent heat $L$ and
the surface tension $\sigma$, we have studied bubble  nucleation and
growth. The effects of a more general parametrization have been
estimated. We have investigated the possibility that the
inhomogeneities generated at the QCD phase transition  significantly
affect nucleosynthesis.

We find that parameter values in the range of, \eg, $\sigma \approx
0.01$--$ 0.1T_c^3$, $L \approx 0.01$--$0.1T_c^4$,  lead to a
transition with large supercooling, relatively large critical bubbles,
and a distance between bubbles of $l_n \approx 10^{-3}$--$10^{-2}
t_H$.  These bubbles may grow as detonations as well as deflagrations,
and the Universe does not reheat to $T_c$. In spite of the large
distance scale of the inhomogeneity, 
phase 
we found that for detonations the accumulation of baryon number in the
high density regions is too ineffective to make a noticeable effect on
nucleosynthesis likely. Even in the new region of parameter space
studied, only deflagrations with efficient turbulent baryon number
transport, or maybe with large neutrino flux,  seem able to affect
nucleosynthesis.


\clearpage

\noindent  FIGURE CAPTIONS
\bigskip  \bigskip

\begin{figure}[h]
\caption{\protect
The supercooling $1-\hat T_f$  and the bubble distance scale $l_n$.
The distance scale is given as $l_n/vt_H$, where $t_H$ is the Hubble
distance (``horizon'') and $v$ is the detonation or shock velocity,
$1/{\protect\sqrt3} < v < 1$. The dependence on the surface tension
$\sigma$ and the latent heat $L$ is through the combination
$\sigma^3/L^2T_c$.  We also show the nucleation action $S(t_f)$.  The
other thin  lines show the two quantities $S''(t_f)/S'(t_f)^2$
(short-dashed line) and $\bigl[|S'(t_f)|(t_f-t_c)\bigr]^{-1}$ (dotted
line), whose smallness we have assumed.  Our approximations are seen
to break down at $\sigma^3/L^2T_c \simgt 0.25$.
}
\label{figB}
\end{figure}

\begin{figure}[h]
\caption{\protect
Contours of the critical radius $r_c$ and the bubble separation $l_n$
on the $(L,\sigma)$ parameter plane.  The solid line corresponds to
$\sigma^3/L^2T_c = 0.25$.  This figure is for the bag equation of
state. 
equation of As discussed in the text, use of a more realistic equation
of state could increase the distances $l_n$ by an order of magnitude
or more.
}
\label{figC}
\end{figure}

\begin{figure}[h]
\caption{\protect
  Schematic representation of the energy density versus $T^4$
for three different equations of state.
Thinner parts of the curves denote
the metastable branches. For clarity the magnitude of $L$
has been exaggerated in the figure.\la{fig:3EOS}}
\end{figure}

\clearpage

\begin{figure}[h]
\caption{\protect
   Detonations and deflagrations:  This plot shows how the different
processes lie in the $(\eph,\epq)$-plane.  The entropy
condition restricts the allowed processes below the $\Delta S = 0$
curve.  Point C corresponds to $T_q = T_h = T_c$.  For a given $T_f$
there is a 1-dimensional family of solutions, denoted by the dashed
line.  The detonation branch of this family is a horizontal line, the
deflagration branch a steep curve.  For any $T_f<T_c$ it always passes
to the left of point C, indicating that weak deflagrations are
allowed.
If it passes to the left of G, strong deflagrations are allowed.
If the detonation branch passes below D (J), then weak (Jouguet)
detonations are allowed. This figure is for $r = 1.01$.
}
\label{eheqfig}
\end{figure}

\begin{figure}[h]
\caption{\protect
   Some special temperatures for the bag equation of state as
a function of $r$.  $\TqD$ is the
maximum temperature for which (weak) detonations are allowed.
$\TqJ$ is the maximum temperature for which Jouguet detonations are
allowed.  $\TfG$ is the maximum temperature for which Jouguet
deflagrations are allowed. For $T<\TfG$, strong deflagrations are
allowed.  $T_r$ is the lowest temperature for which the latent heat is
sufficient to reheat the universe back to $T_c$.
}
\label{tfg}
\end{figure}

\begin{figure}[h]
\caption{\protect
Regions on the $(L,\sigma)$ parameter space, where different processes
are allowed.  This figure is for a bag model with $a_q =
34.25\pi^2/90+L/8T_c^4$, $a_h = 34.25\pi^2/90-L/8T_c^4$. The solid
lines divide the graph in three regions depending on what kind of
detonation bubbles are allowed by hydrodynamic considerations.  Weak
deflagrations are always allowed.  Strong deflagrations are allowed
above the long-dashed line.  The universe reheats to $T_c$ if we are
below the short-dashed line.  Thus detonations are always allowed in
those cases where the universe does not reheat to $T_c$.
}
\label{siglnew}
\end{figure}


\clearpage

\begin{thebibliography}{99}

\bibitem{AHS} J. H. Applegate, C. J. Hogan, and R. J. Scherrer,
Phys. Rev. D 35, 1151 (1987).

\bibitem{GuthWei} A. Guth and E. Weinberg,
Phys. Rev. D 23, 876 (1981).

\bibitem{Kurki-Suonio} H. Kurki-Suonio, Nucl. Phys. B 255, 231 (1985).

\bibitem{qcdpax} Y. Iwasaki {\em et al.} (QCDPAX Collaboration),
Phys. Rev. D 46, 4657 (1992).

\bibitem{grossmann} B. Grossmann and M. L. Laursen,
Nucl. Phys. B 408, 637 (1993).

\bibitem{iwasaki} Y. Iwasaki, K. Kanaya, L. K\"arkk\"ainen,
K.~Rummukainen, and T.~Yoshi\'e,\\
Phys. Rev. D 49, 3540 (1994).

\bibitem{Petersson}
B.~Petersson, Nucl.~Phys.~B (Proc.~Suppl.) 30, 66 (1993).

\bibitem{LL}
L. D. Landau and E. M. Lifshitz, {\sl Fluid Mechanics},
2nd edition \\
(Pergamon Press, Oxford, 1987).

\bibitem{Courant}
R. Courant and K. O. Friedrichs,
{\sl Supersonic flow and shock waves} \\
(Springer-Verlag, Berlin, 1985).

\bibitem{GKKM}
M. Gyulassy, K. Kajantie, H. Kurki-Suonio, and L. McLerran,\\
Nucl. Phys. B 237, 477 (1984).

\bibitem{Steinhardt} P. J. Steinhardt, Phys. Rev. D 25, 2074 (1982).

\bibitem{Laine} M. Laine, Phys. Rev. D, in press.

\bibitem{IKKL}
J. Ignatius, K. Kajantie, H. Kurki-Suonio, and M. Laine,
Phys. Rev. D, in press.

\bibitem{KajKur} K. Kajantie and H. Kurki-Suonio, Phys. Rev. D 34,
1719 (1986).

\bibitem{FMA} G. M. Fuller, G. J. Mathews, and C. R. Alcock,
Phys. Rev. D 37, 1380 (1988).

\bibitem{Kurki-SuonioB} H. Kurki-Suonio, Phys. Rev. D 37, 2104 (1988).

\bibitem{KapOli} J. I. Kapusta and K. A. Olive, Phys. Lett. B 209, 295
(1988).

\bibitem{SKAM} K. Sumiyoshi, T. Kajino, C. R. Alcock,
and G. J. Mathews,
Phys. Rev. D 42, 3963 (1990).

\bibitem{MMAF} G. J. Mathews, B. S. Meyer, C. R. Alcock, and G. M.
Fuller, Astrophys. J. 358, 36 (1990).

\bibitem{AH} J. H. Applegate and C. J. Hogan,
Phys. Rev. D 31, 3037 (1985).

\end{thebibliography}
\end{document}